\documentclass[12pt]{article}
\usepackage{tikz}
\usetikzlibrary{positioning}
\usepackage{amsmath}
\usepackage{amssymb}
\usepackage{graphicx}
\usepackage{algorithm}
\usepackage{algpseudocode}
\usepackage{booktabs}
\usepackage{multicol}
\usepackage{geometry}
\usepackage{multirow}
\usepackage{bm}
\usepackage{xcolor}
\usepackage{hyperref}
\usepackage{natbib}

\begin{document}

\begin{center}
{\Large\bfseries Data-adaptive gene and pathway-based tests for rare-variant associations with survival outcomes}
\end{center}

\bigskip

\begin{center}
Yu Wang$^{1}$, Kwang Woo Ahn$^{1}$, Sarah L. Kerns$^{2}$, William Hall$^{2}$, Petra Seibold$^{3}$, Christopher J. Talbot$^{4}$, Ana Vega$^{5,6,7}$, Barry S. Rosenstein$^{8}$, Nawaid Usmani$^{9}$, Catharine M.L. West$^{10}$, Liv Veldeman$^{11}$, Paul L. Auer$^{1}$, Zhongyuan Chen$^{12}$
\end{center}

\bigskip

\noindent
$^{1}$Division of Biostatistics, Data Science Institute, Medical College of Wisconsin, Wisconsin, USA; \\
$^{2}$Department of Radiation Oncology, Medical College of Wisconsin, Wisconsin, USA; \\
$^{3}$German Cancer Research Center (DKFZ), Heidelberg, Germany; \\
$^{4}$University of Leicester, Leicester, United Kingdom; \\
$^{5}$Grupo Genética en Cáncer y Enfermedades Raras, Instituto de Investigación Sanitaria de Santiago de Compostela, Santiago de Compostela, Spain; \\
$^{6}$Fundación Pública Galega de Medicina Xenómica, Santiago de Compostela, Spain; \\
$^{7}$Biomedical Network on Rare Diseases (CIBERER), Madrid, Spain; \\
$^{8}$Icahn School of Medicine at Mount Sinai, New York, New York, USA; \\
$^{9}$Cross Cancer Institute, University of Alberta, Edmonton, Canada; \\
$^{10}$University of Manchester, Manchester, United Kingdom; \\
$^{11}$Ghent University / Ghent University Hospital, Ghent, Belgium; \\
$^{12}$To be added

\medskip

\noindent
\textbf{Correspondence:} Zhongyuan Chen (zychenstat@gmail.com), Paul Auer (pauer@mcw.edu)

\bigskip

\begin{abstract}
Statistical methods for testing aggregate rare-variant genetic associations are typically based on either burden or dispersion tests (or a combination of the two). These methods lack statistical power in the presence of diverse genetic architectures. Moreover, few aggregate rare-variant association methods have been developed specifically for survival data. To address these issues, we propose data-adaptive gene- and pathway-based association tests based on Schoenfeld residuals in Cox proportional hazards models for association studies between an aggregate of rare-variants and survival outcomes. Our methods improve statistical power while maintaining flexibility across various genetic effect sizes and directions. We develop an efficient R package that enables fast computation and supports data simulation as well as gene- and pathway-level testing. Applying our approach to late bladder toxicity following radiotherapy for non-metastatic prostate cancer, we identify biologically relevant genes and pathways, replicate known signals, and capture additional associations. Our method provides a powerful, adaptive framework for survival-based genetic association studies of rare-variants.
\end{abstract}

\medskip

\noindent\textbf{Keywords:} aSPU, time-to-event outcomes, rare-variant associations, Cox regression, Schoenfeld residuals

\section{Introduction}

\label{sec:sec1}

Over the past two decades, genome-wide association studies (GWAS), which examine associations between traits and single-nucleotide polymorphisms (SNPs) across the genome, have successfully identified thousands of common genetic variants associated with complex human diseases \cite{visscher2017}. However, common variants typically explain only a modest proportion of trait heritability, prompting increased focus on rare variants with potentially larger effect sizes \cite{Lee2014}. Next-generation sequencing technologies now enable systematic investigation of rare and low-frequency variants across the genome \cite{Behjati2013}, yet rare variants pose unique analytical challenges: their low minor allele frequencies provide insufficient power for traditional single-variant tests. Consequently, region- or gene-based association tests that aggregate information across multiple variants within a genomic unit have become essential tools for rare variant analysis \cite{Lee2014}.

Current rare-variant methods fall into several paradigms. \textbf{Burden} tests, such as the combined multivariate and collapsing method, assume that all causal variants within a region influence the trait in the same direction and collapse rare variants into a single genetic score \cite{li2008}. While computationally efficient and statistically powerful under this assumption, burden tests suffer considerable power loss when effects are bidirectional or when many variants in the set are non-causal. Variance-component tests, most notably the sequence kernel association test (\textbf{SKAT}), instead model variant effects as random with mean zero and accommodate heterogeneity in both magnitude and direction \cite{wu2011}. SKAT demonstrates superior power when causal variants include protective and deleterious effects, but may be less efficient than burden tests when most causal variants act in a consistent direction. To address these complementary strengths and weaknesses, combined tests such as SKAT-O optimally combine burden and variance-component approaches \cite{lee2012}. Functional regression approaches offer yet another framework, modeling genetic effects as smooth functions of genomic position rather than assuming fixed or random effects (\textbf{CopulaFLM} \cite{Luo2011, Fan2013}). More recently, copula-based methods have emerged to treat survival outcomes as multivariate, jointly modeling censoring status and observed time through flexible dependence structures (\textbf{CBMAT} \cite{StPierre2022}).

The adaptive sum of powered score (aSPU) test and its pathway extension (aSPUpath) provide flexible omnibus procedures that transcend individual methodological frameworks \cite{Pan2014, Pan2015}. The aSPU family aggregates score-based statistics across variants using varying power parameters, then selects the most informative combination via permutation. This data-adaptive strategy provides robustness across diverse genetic architectures, whether effects are sparse or dense, unidirectional or bidirectional \cite{Pan2014}. However, most aSPU applications have focused on continuous or binary outcomes. Extension to time-to-event data is non-trivial because right-censoring complicates residual construction and score statistic formulation under the Cox proportional hazards model \cite{cox1972regression}.

Despite substantial methodological progress in rare-variant analysis for quantitative and case-control traits, methods for survival outcomes remain comparatively underdeveloped \cite{Lee2014}. Chen et al. extended burden and SKAT-O tests to Cox regression using martingale residuals, demonstrating the feasibility of SNP-set survival analysis \cite{Chen2014}. However, these methods do not incorporate the adaptive, data-driven power-parameter selection that has proven valuable in aSPU-type tests. Moreover, while single-variant time-to-event GWAS tools such as SPACox and the age-dependent liability threshold (ADuLT) model have advanced large-scale survival genomics \cite{SPAcox, ADuLT2023}, they operate at the variant level and do not provide gene- or pathway-level inference for rare variants.

% Functional regression models face challenges when variant positions are not stochastic, and existing extensions primarily address hierarchical correlation structures or bivariate survival outcomes rather than directional adaptivity \cite{Chien2017, Wei2019}. While copula-based approaches provide elegant handling of censored data by modeling the joint distribution of event time and censoring \cite{StPierre2022}, they have not been integrated with adaptive power-parameter selection frameworks like aSPU.

Functional regression models have gained popularity recently. It faces challenges when variant positions are treated as fixed rather than stochastic. For example,  Chien et al. and Wei et al. proposed extensions of functional regression models for survival outcomes \cite{Chien2017, Wei2019}; however, these methods primarily focus on accommodating hierarchical correlation structures or bivariate survival outcomes, rather than enabling directional adaptivity in genetic effect aggregation. St-Pierre et al. also proposed copula-based approaches to address censoring by modeling the joint distribution of event time and censoring \cite{StPierre2022}. While these approaches provide a framework for handling censored data, they have not been integrated with adaptive power-parameter selection frameworks such as aSPU.

The methodological gap is particularly salient for studying late bladder toxicity, including hematuria, following radiotherapy for non-metastatic prostate cancer. Radiotherapy is a primary treatment modality, yet treatment-related side effects significantly impact patient quality of life and treatment satisfaction \cite{Diefenbach2007, Resnick2012}. While common-variant GWAS and polygenic risk score studies have identified genetic contributors to radiotherapy toxicity \cite{Farazi2025}, rare-variant studies remain scarce, in part due to the lack of appropriate statistical methods for analyzing time-to-toxicity outcomes with right censoring. %Note that hematuria usually occurs before death; we do not consider it a competing risk. 

In this work, we extend the aSPU and aSPUpath frameworks to survival outcomes by constructing residual-based score statistics using Schoenfeld residuals from the Cox proportional hazards model \cite{schoenfeld1982, GRAMBSCH1994}. This yields adaptive, directionally sensitive gene- and pathway-level tests for right-censored traits (Section~\ref{s:method}). We evaluate our method through extensive simulations across diverse genetic architectures (Section~\ref{s:sim}) and demonstrate its application to RadioGenomics consortium data for late bladder toxicity following radiotherapy in prostate cancer patients (Section~\ref{s:app}). Our analysis identifies biologically relevant genes and pathways, including signals obscured under traditional burden or variance-component frameworks. An R package \texttt{aSPUS} implementing the proposed methods is publicly available at \href{https://github.com/yuw444/aSPUs}{https://github.com/yuw444/aSPUs}.

\section{Method}
\label{s:method}
\subsection{Notation}

%In apart from the original aSPU method\cite{Pan2014}, 
Let $\pmb{Z}^s_i = (Z_{i1}, \ldots, Z_{iP}) \in \mathbb{R}^{P}$ represent the $p$ SNP genotypes for subject $i$, which are encoded as the number of minor allele counts (0, 1, or 2) and let $\pmb{\beta}_s$ denote the corresponding parameter vector for the Cox model. Note that imputed genotypes (i.e., dosages that take on values on $(0,2)$) can be used for $\pmb{Z}^s_i$\cite{Behjati2013}. We also consider additional covariates $\pmb{Z}^m_i=(Z_{iP+1,\ldots,Z_{iP+K}})^\top \in \mathbb{R}^{K}$ for subject $i$, which may be either fixed or time-dependent. For simplicity, we consider fixed covariates for $\pmb{Z}^m_i$. Let $\pmb{\beta}_m$ be the parameter vector for $\pmb{Z}^m_i$ in the Cox model and define $\pmb{Z}_i^\top=((\pmb{Z}^s_i)^\top,(\pmb{Z}^m_i)^\top)$ and $\pmb{\beta}^\top = (\pmb{\beta}_s^\top, \pmb{\beta}_m^\top)$. 

Let $T_i$ and $C_i$ be event and censoring times, respectively, for subject $i$. We assume $T_i$ and $C_i$ are independent given covariates. Let $X_i=\min(T_i,C_i)$ and $\delta_i$ be the observed time and event indicator for subject $i$, respectively, where $\delta_i = 1$ if the event is observed and 0 otherwise. Denote $H_0(t)$ as a function of the baseline cumulative hazard function at time $t$. The data structure for each patient is shown in the left table of Figure~\ref{fig:perms}. Then, the Cox model is 
\[S(t| \pmb Z_i) = \exp\left\{-H_0(t)\exp(\pmb Z_i^\top \pmb\beta)\right\}, \; i = 1, \ldots, n.\]
% We assume there are $n$ independent subjects. Let $(X_i, \Delta_i)$ denote the observed time and event indicator for subject $i$, where $\delta_i = 1$ if the event is observed and 0 otherwise. Let $\pmb{Z}^m_i \in \mathbb{R}^{K}$ represent the clinical covariates, which may be either fixed or time-dependent, and let $K$ denote the number of covariates. For simplicity, we consider fixed covariates for $\pmb{Z}^m_i$. Additionally, let $\pmb{Z}^s_i = (Z_{i1}, \ldots, Z_{iP}) \in \mathbb{R}^{P}$ represent the SNP information, which may be encoded as the number of minor allele counts (0, 1, or 2), imputed dosage values continuous on $(0,2)$ \cite{Behjati2013}. The data structure for each patient is shown in the left table of Figure~\ref{fig:data}.
Let $Y_l(t) = I(X_l \ge t)$ indicate whether subject $l$ is at risk at time $t$. The score function of the Cox model is 
\[
\pmb U(\pmb \beta) = \sum_{i=1}^n \left\{ \pmb{Z}_i - \frac{\sum_l \pmb{Z}_l e^{\pmb \beta^\top \pmb{Z}_l}Y_l(X_i)}{\sum_l e^{\pmb \beta^\top \pmb{Z}_l}Y_l(X_i)} \right\}\delta_i. 
\]

% , $\pmb{Z}_l = [\pmb{Z}^s_l, \pmb{Z}^m_l]$, and $\pmb{\beta}^T = (\pmb{\beta}_s^\top, \pmb{\beta}_m^\top)$ represents the true effects for SNPs ($\pmb{\beta}_{s}$) and clinical covariates ($\pmb{\beta}_{m}$).

\subsection{Schoenfeld Residuals}
\label{sec:ms}
The original data-adaptive aSPU test\cite{Pan2014} constructs a class of tests by exponentiating the product between the mean residual of the response variable and the covariates. They consider a variety of exponents ($\gamma \in \{1,2, \cdots, 8, \infty\}$) for flexibility. P-values for the original aSPU test are obtained empirically through permutations. In the context of survival outcomes, we consider the Schoenfeld residuals \cite{GRAMBSCH1994} for the proposed test, which is %It is defined by the integrand of (\ref{eq:scores}) as follows:
\begin{align*}
% \label{eq:integrand}
\pmb \epsilon_i(t) = \pmb{Z}_i - \frac{\sum_l \pmb{Z}_l e^{\pmb \beta^\top \pmb{Z}_l}Y_l(t)}{\sum_l e^{\pmb \beta^\top \pmb{Z}_l}Y_l(t)} = \pmb{Z}_i - \frac{\sum_l \pmb{Z}_l e^{\pmb \beta^\top \pmb{Z}_l}I(X_l \ge t)}{\sum_l e^{\pmb \beta^\top \pmb{Z}_l}I(X_l \ge t)}, \ i=1,\ldots,n.
\end{align*}
%We leverage this residual in the proposed permutation test to ease the computing burden.

\subsection{Hypothesis Testing}
In GWAS, investigators are interested in examining the effect of individual SNPs after adjusting for covariates. Here, we consider testing whether the aggregate effect of SNPs within a gene are associated with a survival outcome. This framework will be extended to a pathway-based test in Section~\ref{permutation}. We consider the following hypotheses for a gene-based test:
\[H_0: \pmb{\beta}_s = \pmb{0},\ H_1: \pmb{\beta}_s \ne \pmb{0}.\]
These hypotheses test whether there are SNP effects on the survival outcome after adjusting for covariates. Now, we define
\begin{align}
\label{eq:weight}
\omega_j(t) = \frac{e^{\pmb \beta_m^\top \pmb{Z}^m_j}I(X_j \ge t)}{\sum_l e^{\pmb \beta_m^\top \pmb{Z}^m_l}I(X_l \ge t)},
\end{align}
and observe that (\ref{eq:weight}) is invariant to $\bm Z^s$.
Under $H_0$, we have $\pmb \beta^\top \pmb{Z}_l = \pmb \beta_m^\top \pmb{Z}^m_l$, yielding
\begin{align*}
% \label{eq:epsilon}
  \pmb \epsilon^s_i(t) = \pmb{Z}^s_i - \frac{\sum_l \pmb{Z}^s_l e^{\pmb \beta_m^\top \pmb{Z}^m_l}I(X_l \ge t)}{\sum_l e^{\pmb \beta_m^\top \pmb{Z}^m_l}I(X_l \ge t)} = \pmb{Z}^s_i - \sum_j \pmb{Z}^s_j \omega_j(t).
\end{align*}
%where the second term in (\ref{eq:epsilon}) evaluates the weighted mean of $\pmb{Z}^s$ with weight 
%\begin{align}
%\pmb \epsilon^s_i(t) = \pmb{Z}^s_i - \sum_j \pmb{Z}^s_j \omega_j(t).  
%\end{align}
%Due to the nature of the survival counting process $N_i(t)$, $dN_i(t) = 0$ for any $t \neq x_i$, the event time of subject $i$ with $\delta_i = 1$. Therefore, the integrand $\pmb \epsilon^s_i(t)$ contributes to the summation in (\ref{eq:scores}) only at $x_i$ for subject $i$. For patients observed only at the censoring time point (i.e., $x_i$ with $\delta_i = 0$), the contribution is zero. Thus, (\ref{eq:scores}) simplifies to 
As described in Section~\ref{permutation}, using $\pmb \epsilon^s_i(t)$ greatly improves the computational efficiency of the proposed permutation test. Then, we have the following score function for $\pmb{\beta}_s$ under $H_0$: 
\begin{align}
 \label{eq:scores_i}
\pmb U_s = \sum_{i=1}^n  \left\{ \pmb{Z}^s_i -  \sum_j \pmb{Z}^s_j \omega_j(X_i)\right\}\delta_i.
\end{align}
 % Similar to the aSPU framework, we can then form our testing statistics as follows. 

% \begin{align*}
%     u_\gamma^0 = \left(\sum_{j = 1}^P |\pmb U_{s,j}^{0}|^\gamma\right)^{1/\gamma}, \; \gamma \in N^+. 
%     % \label{eq:gene-obs}
% \end{align*}

%In other words, we only need to calculate $\omega_{ij} = \omega_j(x_i)$ at each observed event time point. Elsewhere, the contribution is zero. We denote $\omega_{i\cdot} = (\omega_{i1}, \omega_{i2}, \cdots, \omega_{in})^T.$

\subsection{Permutation Procedure to Obtain $p$-values}\label{permutation}

Like the existing aSPU methods\cite{Pan2014, Pan2015}, we propose a permutation-based procedure called `aSPUS' to obtain empirical $p$-values as follows. First, consider a gene-level test as described above. Under $H_0$, $\pmb{\beta}_s = \pmb{0}$. So we can permute $\pmb{Z}^s$ repeatedly to obtain the empirical $p$-value through the score function in (\ref{eq:scores_i}) with adaptive exponent $\gamma$.

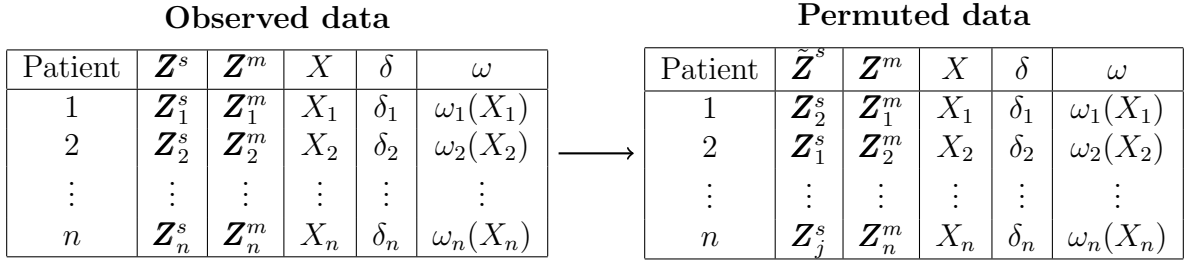
\begin{figure}[!htbp]
    \centering
\begin{center}
\begin{tikzpicture}
    % First table as a node
    \node (table1) [
        label=above:{\textbf{Observed data} }
    ] {
    \begin{tabular}{|c|c|c|c|c|c|c|}
    \hline
    Patient & $\pmb{Z}^s$&  $\pmb{Z}^m$& $X $ & $\delta $ & $\omega$ \\ 
    \hline
    1 & $\pmb{Z}^s_1$& $\pmb{Z}^m_1$& $X_1 $ & $\delta_1 $ & $\omega_{1}(X_1)$\\
    2 & $\pmb{Z}^s_2$& $\pmb{Z}^m_2$& $X_2 $ & $\delta_2 $ & $\omega_{2}(X_2)$ \\
    $\vdots $ & $\vdots$ & $\vdots$ & $\vdots$ & $\vdots $ & $\vdots $ \\
    $n $ & $\pmb{Z}^s_n$& $\pmb{Z}^m_n$& $X_n $ & $\delta_n $ & $\omega_{n}(X_n)$ \\
    \hline
    \end{tabular}
    };

    % Second table as a node, positioned to the right of the first table
    \node (table2) [
        right=of table1,
        label=above:{\textbf{Permuted data} }
    ] {
    \begin{tabular}{|c|c|c|c|c|c|c|}
    \hline
    Patient & $\tilde{\pmb Z}^s$&  $\pmb{Z}^m$& $X $ & $\delta $ & $\omega$ \\ 
    \hline
    1 & $\pmb{Z}^s_2$& $\pmb{Z}^m_1$& $X_1 $ & $\delta_1 $ & $\omega_{1}(X_1)$ \\
    2 & $\pmb{Z}^s_1$& $\pmb{Z}^m_2$& $X_2 $ & $\delta_2 $ & $\omega_{2}(X_2)$  \\
    $\vdots $ & $\vdots$ & $\vdots$ & $\vdots$ & $\vdots $ & $\vdots $ \\
    $n $ & $\pmb{Z}^s_j$& $\pmb{Z}^m_n$& $X_n $ & $\delta_n $ & $\omega_{n}(X_n)$ \\
    \hline
    \end{tabular}
    };
    % Draw an arrow between the tables
    \draw[->, thick] (table1.east) -- (table2.west);
\end{tikzpicture}
\end{center}
\caption{Illustration of $\bm Z^s$ permutation}
\label{fig:perms}
\end{figure}

\begin{figure}[!htbp]
    \centering
    \includegraphics[width=12cm]{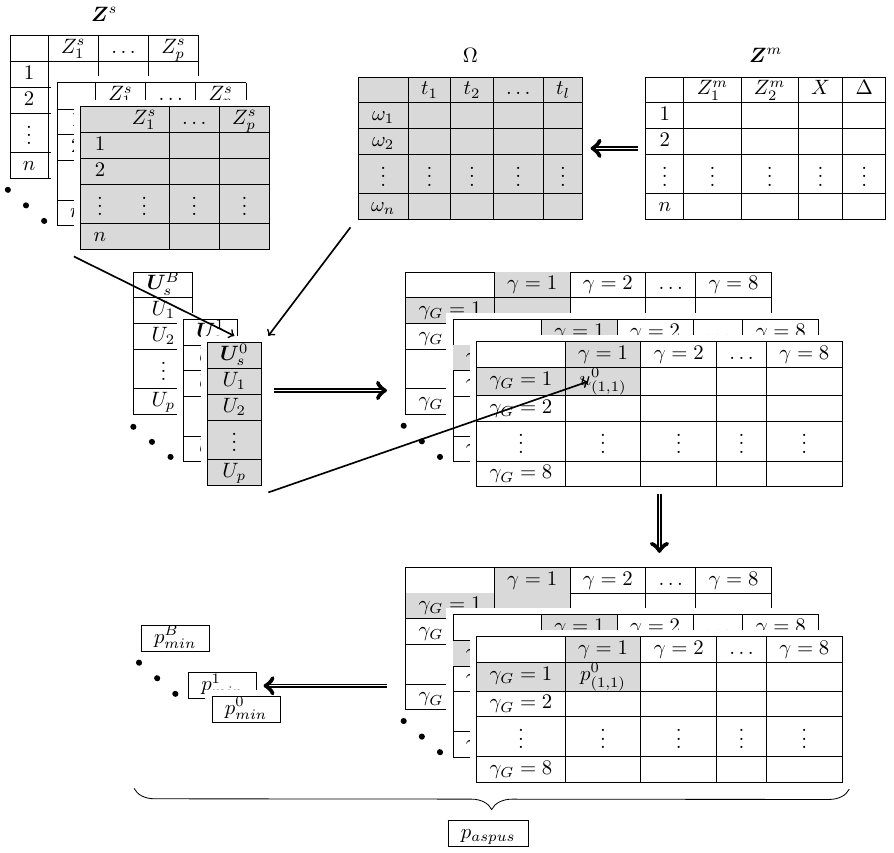}
    \caption{The framework of aSPUS. The top-left stacked tables represent permutations of the variant matrix $\bm Z^s$; the shaded one is observed. The weight matrix $\Omega$ is calculated based on observed times and status for each subject in $\bm Z^m$, remaining constant throughout. For each pair of $(\gamma, \gamma_G)$, combining $\Omega$ with each permutation of $\bm Z^s$ in the score function yields the test statistic $u$. Aggregating $u$ from all permutations produces the corresponding $p$-value. Enumerating all pairs of $(\gamma, \gamma_G)$ finds the minimal $p$-value $p_{min}$ for each permutation. Finally, we compare $p_{min}$ values from permutations and the observation, treating $p_{aSPUS}$ as the final $p$-value for the gene or pathway of interest.}
    \label{fig:framework}
\end{figure}

After permuting $\pmb{Z}^s$ in Figure~\ref{fig:perms}, the position of $(\pmb{Z}^m,X, \delta)$ remains unchanged. As a result, by (\ref{eq:weight}), the value of $\omega_{i}(X_i)$ stays the same after the permutation under $H_0$. This allows the calculation of $\{\omega_1(X_1),\ldots,\omega_n(X_n)\}$ to be done only once for the proposed test, which dramatically improves computing efficiency, as shown in Figure~\ref{fig:framework}. Let $\Tilde{\pmb Z}^{s,b}$ be the permuted ${\pmb Z}^s$ for the $b$th permutation. Then, the score function for the $b$th permutation becomes
\begin{align}
    % \pmb{U} = \sum_{i=1}^n (\pmb z^s_i - \sum_{j = 1}^n \pmb z_i^s \omega_{j}(X_i) ) \delta_i \to 
    \pmb{U}_s^b = \sum_{i=1}^n \left\{\Tilde{\pmb Z}^{s,b}_i - \sum_{j = 1}^n \Tilde{\pmb Z}_i^{s,b} \omega_{j}(X_i) \right\} \delta_i.
    \label{eq:trans}
\end{align}
%Note that $\pmb{U}_s^0$ and $\pmb{U}_s^b$ are vectors of dimension $P \times 1$. 
Assume that we permute ${\pmb Z}^s$ $B$ times. Let $\pmb{U}_s^0$ be the score function based on the original $\pmb{Z}^s_i,i=1\ldots,n$ as in (\ref{eq:scores_i}). 
To obtain the final test statistic, we raise the absolute value of each element to an exponent $\gamma \in \{1, 2, 3, \cdots, \infty\}$, in $\pmb{U}_s^0$ and $\pmb{U}_s^b$, then take the inverse exponent of their summation as follows:
\begin{align}
    u_\gamma^b = \left(\sum_{j = 1}^P |U_{s,j}^{b}|^\gamma\right)^{1/\gamma}, \ b = 0,1,\ldots, B,
    \label{eq:gene-base}
\end{align}
where $U_{s,j}^{b}$ is the $j$th element of $\pmb U_s^b$. In practice, $\gamma \in \{1,2, \cdots, 8, \infty\}$ is used\cite{Pan2014}. As $\gamma$ approaches $\infty$, the $u_\gamma^b$ approaches the maximum element in the vector, i.e., $u_\infty^b = \max\left\{\{U_{s, j}^b, j \in (1, P)\}\right\}$.

As shown in flowchart of Figure~\ref{fig:framework}, for the observation and the $b$th permutations, we calculate one empirical $p$-value $p_\gamma^b$ for each choice of $\gamma$ as follows:
\begin{align*}
    p_\gamma^b = \frac{1}{B+1}\sum_{b'=0}^B I(u^{b'}_\gamma \ge u_\gamma^b), \ b = 0,1,\ldots, B.
\end{align*}
Then, at each permutation, we find the minimal $p$-value with respect to $\gamma$, 
\begin{align*}
    p_{min}^b = \min\left\{p_{\gamma}^b: \gamma \in  \{1,2, \cdots, 8, \infty\} \right\}, \ b = 0,1,\ldots, B.
\end{align*}
Finally, the raw $p$-value of the permutation test is 
\begin{align*}
    p_{aSPUS} = \frac{1}{B+1}\sum_{b=1}^B I\left(p_{min}^0 \le p_{min}^b\right).
\end{align*}

Next, we extend (\ref{eq:gene-base}) to a pathway-level test, where SNPs are located in genes that are themselves located in ''pathways'' or sets of genes. Suppose $P$ SNPs are drawn from a gene set indexed by $\{G_1, G_2, \ldots, G_m\}$, with $k_g$ SNPs in gene $g$ for $g=1,\ldots,m$. Investigators may assign weights at both the SNP and gene levels to reflect prior knowledge about their contributions to the trait. Let $V_s$ denote SNP-level weights $(P \times 1)$ and $Q_g$ denote gene-level weights $(m \times 1)$. The resulting statistic is
\begin{align}
    u_{(\gamma, \gamma_G)}^b =
    \sum_{g = 1}^{m}
    \left(
    q_g
    \left(
    \sum_{s \in G_g}
    (v_s |U_s|)^\gamma / k_g
    \right)^{1/\gamma}
    \right)^{\gamma_G},
    \label{eq:path-based}
\end{align}
where $k_g$ adjusts for the number of SNPs in gene $g$, preventing genes with more SNPs from dominating the statistic. Similar to $\gamma$ to the gene-based test, $\gamma_G$ aggregates the gene-effect at different scale levels in the pathway-based test. The weights $v_s$ and $q_g$ enable the incorporation of prior biological knowledge; for example, genes believed to play a larger role in the pathway may be assigned higher weights. Missense/nonsense SNPs may be given higher weights. When no prior information is available, equal weights can be used. Similar to aSPU framework\cite{Pan2015}, we suggest trying $\gamma \in \{1,2, \cdots, 8\}$ and $\gamma_G \in \{1, 2, 3, \cdots, 8\}$ as shown in the later simulation, though this needs to be further studied. 

Similarly to the permutation procedure used for the gene-level test, we perform permutations for the pathway-level test by replacing the statistic in (\ref{eq:gene-base}) with that in (\ref{eq:path-based}). The resulting empirical distribution is then used to obtain the raw $p$-value for the pathway of interest.

% \textcolor{red}{Please describe how to perform a permutation test here. Citing a previous paper is insufficient.}

\subsection{Pseudo Algorithm of aSPUS}
\label{sec:pseudo}

The proposed aSPUS is computationally intensive, particularly when multiple choices of $\gamma$ and $\gamma_G$ are considered. The burden is further amplified in the time-to-event setting, where $\omega_j(X_i)$ must be evaluated for each patient $j$ at every observed event time. To mitigate these challenges, we modify the original aSPU framework as below to improve computational speed and memory efficiency. Specifically, we have made the following enhancements: 
(i) migrating the core computational routines from \texttt{R} to a $C$ backend via R's C API; 
(ii) indexing permutations of the SNP matrix rather than recreating each permuted matrix, permuting only row indices to reduce memory usage; and 
(iii) implementing a dynamic permutation strategy that begins with a small batch of permutations and adaptively determines whether additional permutations are necessary. Details are provided in the algorithm below.

\begin{itemize}
    \item \textbf{Step 1:} Under $H_0$, that is, SNPs have no effects, fit the Cox model with covariates only and estimate their effects $\bm \beta_m$. 
    \item \textbf{Step 2:} Estimate $\exp({\bm\beta_m^\top \bm Z_i^m})$ and prepare a risk set $\{l: X_l > t\}$ for subject $i=1,\ldots,n$ at all event time points $t$ upfront. Note that these quantities remain invariant when permuting $\pmb{Z}^s$.
    \item \textbf{Step 3:} Calculate $\omega_i(t)$ for subject $i=1,\ldots,n$ at all event time points $t$; these quantities are also invariant to permutation.
    \item \textbf{Step 4:} Calculate $\sum_j\pmb{Z}_j\omega_j(t)$ for each subject at all event time points.
    \item \textbf{Step 5:} Calculate the score $\pmb{U}_s^0$ for the observed data as shown in (\ref{eq:scores_i}).
    \item \textbf{Step 6:} Perform initial $B'$ permutations on $\pmb{Z}^s$, repeat Steps 4--5 to obtain the initial empirical $p$-value.
    \item \textbf{Step 7:} Perform additional $B - B'$ permutations only if the initial $p$-values are below $\theta$.
\end{itemize}
Steps 1 to 5 are performed only once for computational and memory efficiency. Steps 6 and 7 are performed to adaptively determine whether additional $B - B'$ permutations are needed, reducing computational cost when a candidate gene is not significant for the outcome. 
We used $B=500,\; B' = 40$, and $\theta = 0.1$ in this article. More details about the pseudo code are in Algorithm~\ref{alg:pseudo}.

% The original aSPU method described an innovative approach to obtaining adaptive $p$-values for the aSPU test. Here, we  provide a visual representation, as shown in Figure \ref{fig:framework}. The weight matrix $\Omega(n \times d)$, where $d$ is the total number of events, is invariant to permutation of $\bm Z^s$. Combining each permutation of $\tilde{\bm Z^s}$ with $\Omega$, the score statistic $\bm U(P \times 1)$ is obtained. With the choice of $\gamma$ (and $\gamma_G$), the test statistic $U$ is calculated according to (\ref{eq:gene-base}, \ref{eq:path-based}) for each permutation. Then, treating the observation and each permutation as the reference respectively, we calculate the empirical $p$-value for every choice of $\gamma$ (and $\gamma_G$), yielding $|\gamma| (\times |\gamma^G|)$ different $p$-values for each permutation. Finally, the minimal $p$-value is identified for the observation ($p_{min}^o$) and each permutation ($p_{min}^b$, $b \in \{1, 2, \cdots, B\}$), respectively. The final $p$-value of aSPU is calculated as:
% \[p_{aSPUS} = \frac{1}{B+1}\sum_{k=0}^B I(p_{min}^o < p_{min}^k)\]

\section{Simulation}
\label{s:sim}
\subsection{Data Generation}

Burgess et al.~\cite{Burgess_2015} investigated correlated instrumental variables in Mendelian randomization by generating target correlations among SNPs using a Wishart distribution. Adapting their strategy, we extend the algorithm to the time-to-event setting. Specifically, the SNP matrix ($\pmb{Z}_p^s$), the covariate matrix ($\pmb{Z}_K^m$), and the time-to-event outcomes are generated according to the following procedure.

\begin{equation*}
\begin{aligned}
    \Lambda &\sim \text{Wishart}(P, \Lambda_0), \qquad \Phi = \text{Cor}(\Lambda),  \\ 
    \boldsymbol{\Psi}_{1i}, \boldsymbol{\Psi}_{2i} &\overset{\text{iid}}{\sim} \mathcal{N}_P(\mathbf{0}, \Phi), \quad i = 1,\ldots,n, \\
    z_{ip} &= \mathbf{1}_{\Psi_{1ip} > \pi_p} + \mathbf{1}_{\Psi_{2ip} > \pi_p}, 
    \quad \pi_p = \text{qnorm}(f_p), \quad p=1,\ldots,P, \\
    z_{ik} &\sim N(0,1), \quad k=P+1,\ldots,P+K, \\
    % \beta_0 &= \log\left(\frac{a_0}{1-a_0}\right), \\
    \beta_p &\sim \mathrm{Uniform}(0.5a, 1.5a) \cdot \mathrm{Rademacher}(0.5), \quad p=1,\ldots,P, \\
    \beta_k &= 0 \text{ or } 0.1, \quad k=P+1,\ldots,P+K, \\
    T &\sim S(t \mid \mathbf{Z})=\exp\left\{-H_0(t)\exp(\mathbf{Z}^\top\pmb{\beta})\right\}, \\
    C &\sim \text{Uniform}(0,\tau), \\
    X &= \min(T,C), \\
    \Delta &= \mathbf{1}_{T\le C}.
\end{aligned}
\end{equation*}
First, the SNP matrix is generated from two independent multivariate normal vectors $\boldsymbol{\Psi}_{1i}$ and $\boldsymbol{\Psi}_{2i}$ with correlation matrix $\Phi$. To introduce realistic linkage disequilibrium (LD) patterns, the covariance matrix $\Lambda$ is sampled from a Wishart distribution with target correlation matrix $\Lambda_0$, and the correlation matrix $\Phi$ is derived from $\Lambda$.

Biologically, the vectors $\boldsymbol{\Psi}_{1i}$ and $\boldsymbol{\Psi}_{2i}$ represent the two haplotypes of individual $i$. For each SNP $p$, a desired minor allele frequency  $f_p$ is specified. The threshold $\pi_p = \text{qnorm}(f_p)$ is chosen such that the expected allele count approximately satisfies
\[
\sum_i \mathbf{1}_{\Psi_{1ip} > \pi_p} \approx n f_p,
\]
ensuring that simulated variants fall within the specified minor allele frequency range.

In addition, $K$ covariates are generated from a standard normal distribution. SNP effect sizes are drawn from $\text{Uniform}(0.5a,1.5a)$, where $b$ represents the average effect size among causal SNPs. The sign of each SNP effect is randomly assigned using a Bernoulli distribution, allowing variants to have either risk-increasing or protective effects. The effects of the covariates, $\beta_k$, are set to either $0$ or $0.1$.

Survival times are generated from a Cox proportional hazards model with cumulative baseline hazard $H_0(t)$ following a $\text{Weibull}(1,1)$ distribution. The event time $T$ is sampled from the corresponding survival function defined by the baseline hazard and the linear predictor $\mathbf{Z}^\top\pmb{\beta}$. The censoring time $C$ is generated from $\text{Uniform}(0,\tau)$, where $\tau$ is chosen to yield an approximate event rate of $60\%$.

\begin{table}[htbp]
\centering
\begin{tabular}{|c|c|c|c|}
\hline
Scenario & $P$ (SNPs) & $\boldsymbol{\Psi}$ Correlation & Causal SNPs in $Z$ \\
\hline
1 & Gene-based & Independent & Retained \\
2 & Gene-based & Correlated & Retained \\
3 & Gene-based & Correlated & Removed \\
4 & Pathway-based & Independent & Retained \\
5 & Pathway-based & Correlated & Retained \\
6 & Pathway-based & Correlated & Removed \\
\hline
\end{tabular}
\caption{Simulation scenarios considered in the study.}
\label{tab:simulation-scenarios}
\end{table}

We evaluate six simulation scenarios summarized in Table~\ref{tab:simulation-scenarios}. For both gene-based and pathway-based analyses, three SNP correlation structures are considered:
(i) independent SNPs;
(ii) correlated SNPs generated with $\Lambda_0=\text{diag}(0.8)$; and
(iii) correlated SNPs generated with $\Lambda_0=\text{diag}(0.8)$, where causal SNPs are removed from the observed SNP matrix $\pmb{Z}_s$, representing situations in which causal variants are unobserved due to technological limitations.

Under each scenario, we investigate the relationship between statistical power and several factors, including the effect size, the number of causal SNPs within a gene (or pathway), and the total number of SNPs within a gene (or pathway), as well as the control of Type I error at the nominal significance level 0.05. When evaluating Type I error, we set $\pmb{\beta}_s =(\beta_1,\ldots,P)^\top = \pmb{0}$ for each scenario. For gene-based tests, we consider $\gamma \in \{1,2,4,8,\infty\}$, while for pathway-based tests we consider $\gamma \in \{1,2,4,8\}$ and $\gamma_G \in \{1,2,4,8\}$.

For gene-based scenarios, the number of causal SNPs ({\it n\_causal\_snps}) is set to $\{1,3,5\}$, the total number of SNPs ({\it n\_snps}) is $\{10,20,50\}$, and the effect size $a\in\{0,0.2,0.3,0.4,0.5,0.6\}$. For pathway-based scenarios, each gene contains at most one causal SNP. Each pathway consists of 20 genes, and the number of SNPs per gene is sampled from either $\text{Uniform}(2,20)$ or $\text{Uniform}(3,100)$, resulting in approximately 200 or 1000 SNPs in total.

\begin{figure}[!htbp]
\centering
\includegraphics[width=\linewidth]{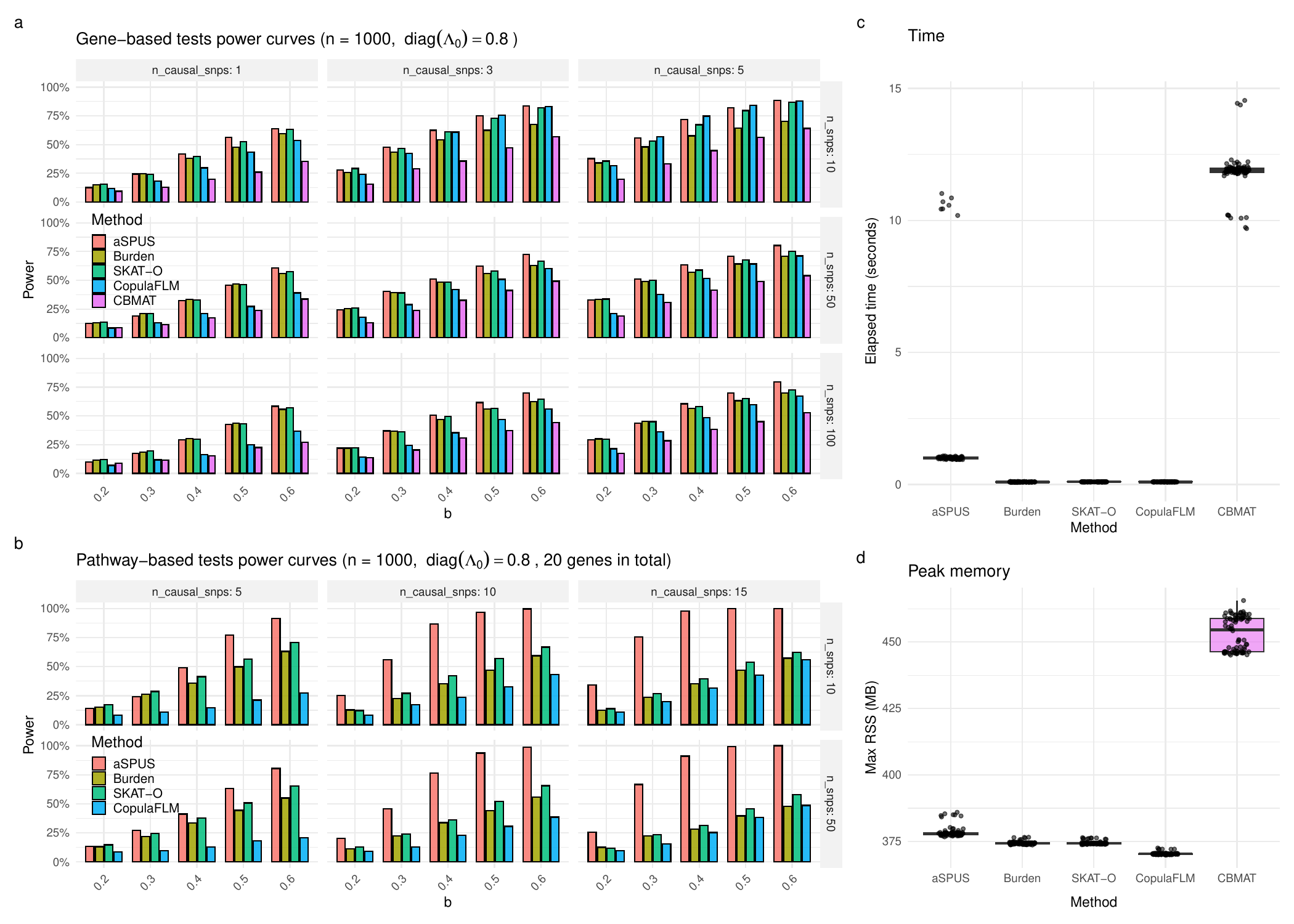}
\caption{Simulation results for gene- and pathway-based tests comparing three methods. \textbf{a} Power versus effect size $\beta\in(0,0.6]$ for combinations of 1, 3, and 5 causal SNPs in genes containing 10, 50, or 100 SNPs under a correlated SNP structure. Both aSPUS and the Burden test gain power as effect size increases, while aSPUS performs slightly better when genes contain more SNPs. \textbf{b} Power versus effect size $\beta\in(0,0.6]$ for pathways consisting of 20 genes with 5, 10, or 15 causal genes and 10 or 50 SNPs per gene. In both settings, aSPUS demonstrates a consistent advantage over the other methods. \textbf{c--d} Computational benchmarks based on 10 simulation replicates with 2,000 subjects and 80 SNPs in gene-based tests. Although aSPUS requires more CPU time, its memory usage is comparable to the other parametric methods.}
\label{fig:rst_sim}
\end{figure}

Across all scenarios, the minor allele frequency $f_p$ is sampled from $\text{Uniform}(0.001,0.05)$, and the baseline hazard follows $H_0(t)\sim \text{Weibull}(1,1)$. The background disease prevalence is fixed at $a_0=0.05$. In the simulations, we compare the proposed method with the traditional Burden test, SKAT-O, CopulaFLM, and CBMAT tests introduced in Section~\ref{sec:sec1}.  For each configuration, we simulate 1,000 observations per dataset, perform 500 permutations for the \texttt{aSPUS} test, and repeat the experiment over 500 simulation replicates to estimate statistical power and Type I error.
% \textcolor{red}{Need to list competing methods. what's the sample size? what's the censoring rate? what's the monte carlo simulation size?}

\subsection{Simulation Results}

Figure~\ref{fig:rst_sim}a, b show the results on power under scenarios 2 and 4, where SNPs are correlated and the causal SNPs are retained. In gene-based tests, the five models show very similar overall trends. Detection power increases with effect size. Given a fixed total number of SNPs, power also increases with the number of causal SNPs. Conversely, fixing the number of causal SNPs, detection power decreases as the total number of SNPs increases. CBMAT and CopulaFLM demonstrate the lowest power in these simulations, highlighting the challenge of detecting signal in the time-to-event setting. The Burden test has similar power to aSPUS and SKAT-O in the presence of a single causal SNP. But the power of the Burden test compared to aSPUS and SKAT-O noticeably drops with greater than 1 causal variant. This observation is expected as we simulated causal variants to have heterogeneous effects that could have opposite signs. The impact of heterogeneous causal effects is even more noticeable when SNPs are independent, as in Figure~\ref{fig:supp_power}a. On the other hand, aSPUS has similar performance to SKAT-O, if not better (Figure~\ref{fig:rst_sim}a). Particularly as the proportion of causal SNPs to the total number of SNPs decreases, aSPUS demonstrates a clear power advantage, suggesting that aSPUS is better at detecting sparse signals. This improvement becomes even more noticeable when SNPs are independent, as shown in Figure~\ref{fig:supp_power}a. In Figure~\ref{fig:supp_power}b, where causal SNPs are removed, aSPUS, Burden, and SKAT-O all show similar power.

% In gene-based tests, the five models show very similar overall trends as in Figure \ref{fig:rst_sim}a. Detection power increases with effect size. Given a fixed total number of SNPs, power also increases with the number of causal SNPs. Conversely, fixing the number of causal SNPs, detection power decreases as the total number of SNPs increases. The burden test shows low power compared to SKAT-O and the proposed framework, where the gene has more than 1 causal SNPs, as expected under heterogeneous effect directions—aggregating opposite effects weakens detection power. In most cases, aSPUS performs similarly to (if not better than) survival SKAT-O. Since aSPUS can adaptively choose exponents on the residuals. In general, we observe aSPUS achieves a better power when the ratio, between the number of SNPs and causal SNPs within the gene, is low(1:5, 1:50, 3:50, 5:50, . It shows a small advantage when the number of causal SNPs is 1 and SNPs are independent. Even in the absence of causal SNPs, tests can still detect some signal when SNP correlation is high, albeit with slightly lower power. Type I error is well-controlled across all scenarios; aSPUS has a slightly higher value but remains close to 0.05 (Table~\ref{tab:alpha_gene}).

In pathway-based tests, we observe a similar overall trend (Figure \ref{fig:rst_sim}b). (\textbf{Yu:  please relabel Figure3b to n.causal.genes instead of n.causal.snps}) Due to the long run time of CBMAT with multiple genes and SNPs, it was excluded from pathway-based tests. For all methods, power increased with effect size and the number of causal genes but the distinction among the four methods becomes more prominent in pathways. The advantages of aSPUS are more evident regardless of whether SNP correlation is independent, correlated, or causal-removed-correlated. (\textbf{Yu: refer to Figure 3b and A1}) As we simulated one causal SNP within each causal gene, the ratio between $n\_causal\_snps$ and $n\_snps$ is fixed around 1/10 and 1/50 for the two $n\_snps = 10, 50$ settings. As was shown for gene-level tests, aSPUS is also most powerful at detecting sparse signals in pathways (\textbf{Yu: refer to both Figures again}).   

Table~\ref{tab:alpha_gene} and ~\ref{tab:alpha_path} summarize the results of the Type I error simulations. In general, aSPUS controls the Type I error at approximately 0.05, with a minimum of 0.0440 and a maximum of 0.0580. CBMAT demonstrates inflated Type I error in gene-based tests. CopulaFLM, Burden, and SKAT-O show reasonable control of Type I error in both gene-based and pathway-based settings.  

Besides power and Type I error, we benchmarked the five methods for computational time and memory consumption. Although Equation \ref{eq:weight} shows that $\omega$ is invariant to permutation under the null hypothesis, repeated computation of $\pmb U_s^b$ and $u^b_{\gamma, \gamma_G}$ for each combination of $\{\gamma, \gamma_G\}$ across permutations imposes a substantial computational burden. Consequently, aSPUS requires longer runtimes and greater memory usage in exchange for potential power gains. To mitigate these costs, we implemented several optimization strategies besides the acceleration in Equation~\ref{eq:trans}. The core procedures (Steps 2--6 in Section \ref{sec:pseudo}) were coded in C to accelerate for-loops, and rather than copying the entire permuted covariate matrix during each iteration, only permuted row indices were stored, reducing memory overhead. As shown in Figures \ref{fig:rst_sim}c and \ref{fig:rst_sim}d, aSPUS requires more computation time but demonstrates comparable memory usage relative to the other methods.

All benchmarks were conducted on a single core and thread of an Intel(R) Xeon(R) Gold 6240R CPU @ 2.40 GHz, using 2,000 subjects, 80 SNPs, and 500 permutations. Reported CPU time and memory usage represent means of 10 replicated simulations. Leveraging modern CPU parallelism, aSPUS can feasibly scale to thousands of genes or pathways.

\section{Application}
\label{s:app}
To demonstrate the practical benefits of aSPUS, we analyzed genetic data from a cohort of men with non-metatstic prostate cancer that have been used to study late bladder toxicity following radiotherapy \cite{Farazi2025}. The primary outcome is time to patient-reported gross hematuria [grade $\ge 2$ ($\ge$G2)]. Following preprocessing, quality control, and imputation procedures described in \cite{Farazi2025}, the analytic dataset comprises 5,997 patients and 11,766 SNPs mapped to 6,910 genes, with the frequency of the number of SNPs within each gene summarized in Table~\ref{tab:snps_per_gene}. Population structure was estimated using principal components analysis (PCA), and the top 10 PCs along with the categorical variable that represents different hospitals, were incorporated into the covariate matrix $\bm Z_m$ in the proposed model.

Both CopulaFLM and CBMAT software cannot analyze genes with a single SNP, so we skip such genes and return NA in the gene-based approach. Additionally, the CopulaFLM framework cannot readily reflect pathways spanning multiple chromosomes, as it requires position as a covariate and ignores chromosome origin. CBMAT suffers from long runtime, especially given the much larger sample size than in simulations. Therefore, we excluded both CopulaFLM and CBMAT from pathway-based analyses.

\begin{figure}[!htbp]
    \centering
    \includegraphics[width=0.8\linewidth]{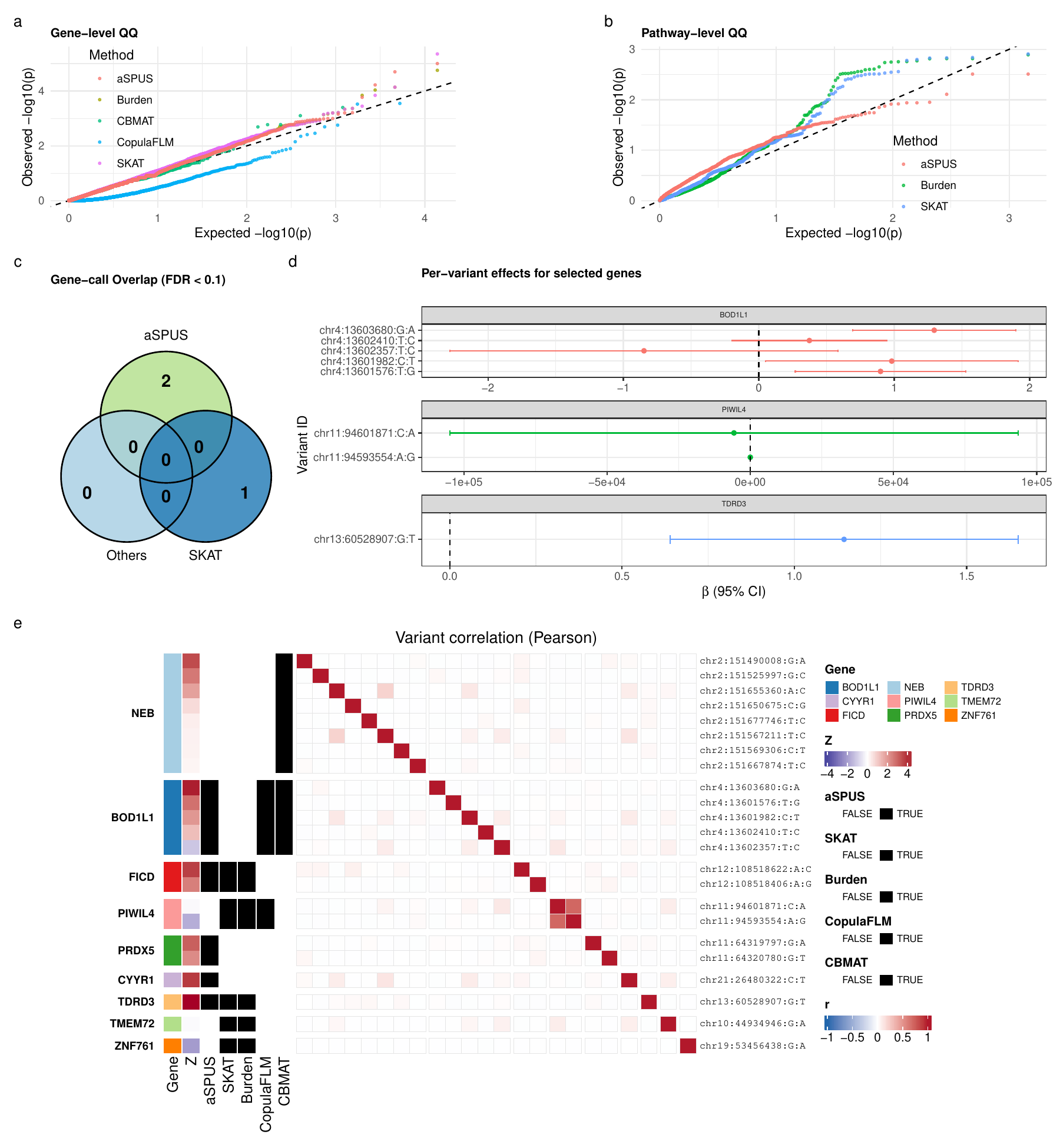}
    \caption{Summary of Data Analysis. \textbf{a} and \textbf{b} show QQ plots at gene- and pathway-levels. In gene-level tests, all methods except CopulaFLM have observed and expected $p$-values well-aligned along the diagonal. \textbf{c} shows detected gene-call overlap among the five methods, with FDR < 0.1. \textbf{d} Decomposes genes with per-variant adjusted effects for detected genes. \textbf{e} Summarizes the top 5 genes detected by at least one of the five methods with $\alpha = 0.05$. The central heatmap is colored by correlation between variant dosage. Genes are grouped and colored with their variants, ordered by $Z$ score from fitting Cox regression with the variant alone. Left panel columns indicate whether each method calls this gene.}
    \label{fig:data}
\end{figure}

For gene-based tests, in addition to the five methods mentioned above, we fitted Cox regression \cite{cox1972regression} with each variant within a gene along with other covariates, obtaining effect size, confidence interval (CI), and $Z$ score for each variant. For pathway-based tests, we first filtered pathways as follows: we ran a Gene Set Enrichment Analysis \cite{Wu2021} against the 6,910 gene-based tests with KEGG, GO, HALLMARK, and Reactome databases, after controlling for multiple testing with $q$-value cutoff of 0.2 using the Benjamini–Hochberg (BH) $p$-value adjustment; 726 pathways passed this filter. We then ran aSPUS, Burden, and SKAT-O on the 726 pathways. 

In gene-based tests, all methods are well-calibrated except CopulaFLM, which tends to overestimate $p$-values (Figure~\ref{fig:data}a). At the pathway-level, both Burden and SKAT-O tests tend to underestimate $p$-values, showing tails above the diagonal line (Figure~\ref{fig:data}b).

With FDR < 0.1, SKAT-O identified 1 gene (PIWIL4), aSPUS identified 2 genes (BOD1L1 and TDRD3), and no genes were identified by other tests; no genes are shared among methods (Figure~\ref{fig:data}c). For the identified genes, adjusted per-variant effects are plotted by gene in Figure~\ref{fig:data}d. In particular, \textit{BOD1L1} contains 4 variants with both protective and adverse survival effects, which is difficult for the Burden test to detect. The SNPs from \textit{PIWIL4} have large variance, so they were detected by the SKAT-O variance test.

In Figure~\ref{fig:data}e, we explore common ground among the five methods and summarize the top 5 genes detected by at least one method at the nominal significance level $\alpha = 0.05$. SKAT-O and Burden share an identical set of genes; both CopulaFLM and CBMAT support the significance of \textit{BOD1L1}. We examined pairwise correlation among variants in detected gene calls to prevent linkage disequilibrium (LD) confounding, finding minimal LD present in detected genes, mainly affecting \textit{PIWIL4}. Genes are grouped and colored with variants, ordered by $Z$ score from per-variant adjusted Cox regression, with exact variants annotated on the heatmap for easy reference. From a biological perspective, all these genes are either validated or potential prognostic markers for certain cancer types \cite{Uhln2015, Uhlen2017}, primarily renal-related cancers (Table~\ref{tab:cancer}). \textit{BOD1L1} has been previously reported as significantly associated with late-bladder toxicity in this cohort, from a Burden-type test \cite{Farazi2025}. 

Finally, no pathways remained significant with FDR < 0.1. For completeness, we report pathways with raw $p < 0.05$ in supplementary files.

\begin{figure}[!htbp]
    \centering
    \includegraphics[width=0.3\linewidth]{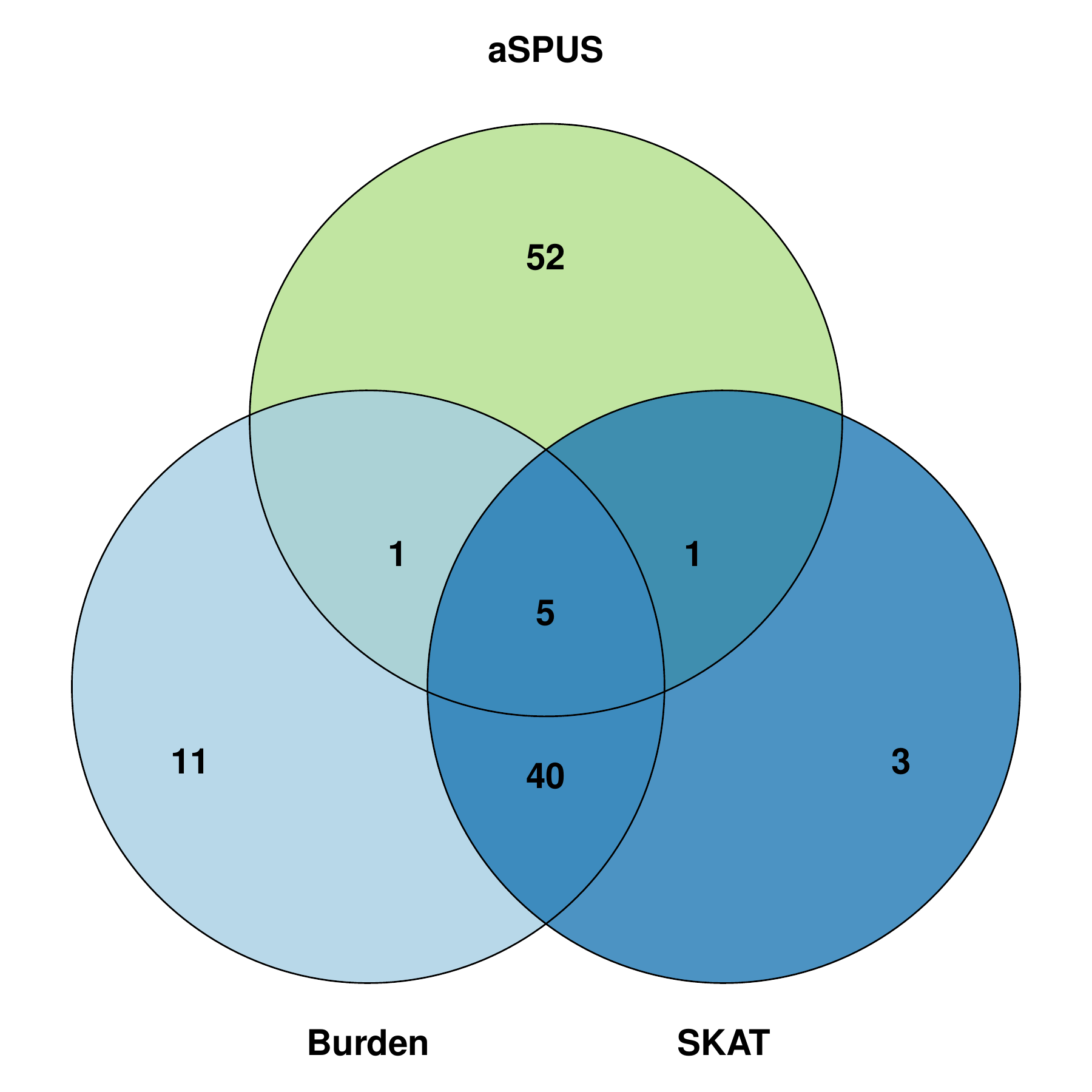}
    \caption{Pathway-call overlap among aSPUS, Burden, and SKAT, with the filter raw $p$-value < 0.05}
    \label{fig:pathway_venn}
\end{figure}
\section{Discussion}

We have proposed an extension of the aSPU test for gene- and pathway-level tests of genetic associations with survival outcomes using Schoenfeld residuals along with an accompanying R package, \texttt{aSPUS} (\href{https://github.com/yuw444/aSPUS}{github.com/yuw444/aSPUS}). Our implementation enables fast computation through a C-based backend and supports data simulation as well as gene- and pathway-level testing. Compared to existing approaches, the proposed method demonstrates particular advantages in scenarios where signals vary in direction and sparsity, especially in pathway-based analyses where two levels of adaptive aggregation are applied: (i) summing over SNPs within each gene and (ii) summing over genes within a pathway. Although the framework can incorporate SNP-level and gene-level weights, our simulation studies evaluated the unweighted case to focus on core operating characteristics. 

In the data application, we have applied aSPUS to a radiogenomics cohort investigating the genetic architecture of late bladder toxicity (grade $\ge 2$ hematuria) following radiotherapy for non-metastatic prostate cancer. While no pathways reached significance under FDR control, several nominally enriched pathways converged on biologically coherent themes—including epithelial polarity, cilia/microtubule signaling, and lipid/inflammatory metabolism—that may contribute to individual variability in late radiation response. These findings highlight aSPUS's utility in detecting weak, multi-gene signals aligning with plausible biological mechanisms, even in settings with modest sample sizes or diffuse genetic effects.

Several directions remain for future work. First, the proposed aSPUS is readily applicable to the proportional cause-specific hazards model for competing risks data. An extensive simulation study under the competing risks settings warrants further investigation. Second, the current version does not incorporate kinship or relatedness among subjects. Integrating a kinship matrix or random-effects structure would allow aSPUS to be applied more robustly in biobank-scale datasets and family-based studies. Finally, exploring adaptive weighting strategies—such as functional annotations, minor allele frequency-based weights, or gene-level priors—may further enhance power, particularly in pathway-level tests where biological structure can be leveraged.

Overall, aSPUS provides a flexible and computationally efficient framework for gene- and pathway-level association testing in survival analyses, with promising performance in both simulation and real-data applications. Its extensibility positions it as a useful tool for emerging genomic and large-scale time-to-event studies.

%\backmatter
\section*{Author contributions}

Z.C., Y.W., P.A., and K.W.A. conceived the idea(s); Y.W. developed the package and conducted the simulations; S.K., W.H., P.S., C.T., A.V., B.R., N.U., C.W., and L.V. performed data curation; Y.W. analyzed the results; Y.W., K.W.A., P.A., and Z.C. wrote the manuscript; and all authors reviewed the manuscript.

\section*{Acknowledgments}
This research was supported in part by {the U.S. National Cancer Institute (U24CA076518) and the U.S. Health Resources and Services Administration (HHSH25020170006C)}. Funding for A. Vega was provided by ISCIII, an initiative of the Ministerio de Ciencia e Innovación, partially supported by European Regional Development FEDER Funds (PI25/00744, DTS24/00083, PI22/00589, INT24/00023), with additional support from the Xunta de Galicia (Programa de Consolidación y Estructuración de Unidades de Investigación Competitivas) GAIN (GPC-IN607B2025/09), the Centro de Investigación Biomédica en Red de Enferemdades Raras CIBERER (ACCI 2016: ER17P1AC7112/2018); the Mutua Madrileña Foundation (FMMA) and Asociación Española Contra el Cáncer (AECC) This research was completed in part with computational resources and technical support provided by the Research Computing Center at the Medical College of Wisconsin.

\section*{Financial disclosure}

None reported.

\section*{Conflict of interest}

The authors declare no potential conflict of interests.

\bibliographystyle{plainnat}
\bibliography{reference}

\section*{Supporting information}

% Additional supporting information may be found in the
% online version of the article at the publisher's website.
The supplement materials are provided.
\newpage
\appendix

% Reset counters for appendix and use "A" prefix
\makeatletter
\counterwithin{figure}{section}
\counterwithin{table}{section}
\makeatother

\section{Program pseudo codes\label{app1}}
\begin{algorithm}[!htbp]
\caption{aSPUS algorithm for survival data}
\label{alg:pseudo}
\tiny
\begin{algorithmic}[1]

\State \textbf{Step 1: Fit Cox model on clinical covariates}
\State Estimate $\hat{\boldsymbol{\beta}}_m$ from the Cox model fitted with clinical covariates

\Statex
\State \textbf{Step 2: Construct the risk-set indicator matrix}
\State Compute $\boldsymbol{\mu}^m = (\mu_1^m, \mu_2^m, \dots, \mu_n^m)^\top$, where
\[
\mu_i^m = \exp\!\left( \hat{\boldsymbol{\beta}}_m^\top \boldsymbol{Z}_i^m \right), \quad i = 1, \dots, n,
\]
and define $\boldsymbol{A} = (a_{ij})$ with
\[
a_{ij} = I(x_j \ge x_i), \quad i,j = 1, \dots, n.
\]

\For{$i = 1$ to $n$}
    \State $u_i \gets \exp\!\left( \hat{\boldsymbol{\beta}}_m^\top \boldsymbol{Z}_i^m \right)$
    \For{$j = 1$ to $n$}
        \State $a_{ij} \gets I(x_j \ge x_i)$
    \EndFor
\EndFor

\Statex
\State \textbf{Step 3: Compute the weight of patient $j$ at time $x_i$}
\For{$i = 1$ to $n$}
    \State $o_i \gets \boldsymbol{A}[i, ] \cdot \boldsymbol{\mu}^m$
    \For{$j = 1$ to $n$}
        \State $\omega_{ij} \gets \dfrac{\mu_j^m \, a_{ij}}{o_i}$
    \EndFor
\EndFor

\Statex
\State \textbf{Step 4: Calculate the weighted mean $\bar{\boldsymbol{z}}^s(x_i)$ for all event times}
\For{$i = 1$ to $n$}
    \If{$\delta_i = 1$}
        \State $\bar{\boldsymbol{z}}^s(x_i) \gets \boldsymbol{0}$
        \For{$j = 1$ to $n$}
            \State $\bar{\boldsymbol{z}}^s(x_i) \gets \bar{\boldsymbol{z}}^s(x_i) + \boldsymbol{z}_j^s \omega_{ij}$
        \EndFor
    \EndIf
\EndFor

\Statex
\State \textbf{Step 5: Calculate the observed score $\boldsymbol{U}^s$}
\State $\boldsymbol{U}^s \gets \boldsymbol{0}$
\For{$i = 1$ to $n$}
    \If{$\delta_i = 1$}
        \State $\boldsymbol{U}^s \gets \boldsymbol{U}^s + \left( \boldsymbol{z}_i^s - \bar{\boldsymbol{z}}^s(x_i) \right)$
    \EndIf
\EndFor

\Statex
\State \textbf{Step 6: Initial permutation procedure}
\For{$\mathrm{perm} = 1$ to $40$}
    \State Permute $\boldsymbol{Z}^s$
    \State Repeat Steps 4--5 to obtain $\boldsymbol{U}^{\prime s}$
\EndFor
\State Obtain the initial empirical $p$-value using the aSPU framework

\Statex
\State \textbf{Step 7: Additional permutations as needed}
\If{$p\text{-value} < 0.1$}
    \For{$\mathrm{perm} = 41$ to $nPerms$}
        \State Permute $\boldsymbol{Z}^s$
        \State Repeat Steps 4--5 to obtain $\boldsymbol{U}^{\prime s}$
    \EndFor
    \State \Return final empirical $p$-value using the aSPU framework
\Else
    \State \Return initial $p$-value
\EndIf

\end{algorithmic}
\end{algorithm}
\normalsize
\newpage
% \section{QQ plots from Application\label{app1}}
% \begin{figure}[b]
%     \centering
%     \includegraphics[width=0.8\linewidth]{fig/qq_plots_gene_10k.pdf}
%     \caption{QQ Plots for Three Methods, aSPUS, Burden, and SKAT-O in Gene-based test}
%     \label{fig:gene_qq}
% \end{figure}

% \begin{figure}[b]
%     \centering
%     \includegraphics[width=0.8\linewidth]{fig/qq_plots_pathway_100k.pdf}
%     \caption{QQ Plots for Three Methods, aSPUS, Burden, and SKAT-O in Pathway-based test}
%     \label{fig:path_qq}
% \end{figure}

\begin{table}[!htbp]
\centering
\caption{Type I error rates by method, correlation structure, and number of SNPs in gene-based test}
\label{tab:alpha_gene}
\begin{tabular}{lccccccccc}
\hline
\multirow{2}{*}{\textbf{Method}} & 
\multicolumn{3}{c}{\textbf{Independent}} & 
\multicolumn{3}{c}{\textbf{Correlated}} & 
\multicolumn{3}{c}{\textbf{Correlate\_removed}} \\
 & \textbf{10 SNPs} & \textbf{50 SNPs} & \textbf{100 SNPs} & 
   \textbf{10 SNPs} & \textbf{50 SNPs} & \textbf{100 SNPs} & 
   \textbf{10 SNPs} & \textbf{50 SNPs} & \textbf{100 SNPs} \\
\hline
aSPUS        
& 0.0513 & 0.0500 & 0.0447 
& 0.0473 & 0.0447 & 0.0580 
& 0.0473 & 0.0447 & 0.0580 \\

Burden       
& 0.0560 & 0.0527 & 0.0467 
& 0.0393 & 0.0433 & 0.0553 
& 0.0440 & 0.0433 & 0.0553 \\

SKAT-O         
& 0.0487 & 0.0440 & 0.0440 
& 0.0433 & 0.0460 & 0.0553 
& 0.0467 & 0.0447 & 0.0553 \\

CopulaFLM    
& 0.0580 & 0.0527 & 0.0460 
& 0.0553 & 0.0480 & 0.0580 
& 0.0553 & 0.0480 & 0.0580 \\

CBMAT        
& 0.0587 & 0.0987 & 0.1107 
& 0.0653 & 0.0634 & 0.0650 
& 0.0653 & 0.0634 & 0.0650 \\
\hline
\end{tabular}
\end{table}

\begin{table}[!htbp]
\centering
\caption{Type I error rates by method, correlation structure, and number of SNPs in pathway-based test}
\label{tab:alpha_path}
\begin{tabular}{lcccccc}
\hline
\multirow{2}{*}{\textbf{Method}} & 
\multicolumn{2}{c}{\textbf{Independent}} & 
\multicolumn{2}{c}{\textbf{Correlated}} & 
\multicolumn{2}{c}{\textbf{Correlate\_removed}} \\
 & \textbf{10 SNPs} & \textbf{50 SNPs} 
 & \textbf{10 SNPs} & \textbf{50 SNPs} 
 & \textbf{10 SNPs} & \textbf{50 SNPs} \\
\hline
aSPUS        
& 0.0440 & 0.0440 
& 0.0440 & 0.0520 
& 0.0440 & 0.0520 \\

Burden       
& 0.0500 & 0.0640 
& 0.0520 & 0.0580 
& 0.0520 & 0.0580 \\

SKAT-O         
& 0.0500 & 0.0460 
& 0.0480 & 0.0480 
& 0.0480 & 0.0480 \\

CopulaFLM    
& 0.0480 & 0.0640 
& 0.0420 & 0.0600 
& 0.0420 & 0.0600 \\
\hline
\end{tabular}
\end{table}

\begin{table}[!htbp]
centering
\caption{Distribution of the number of SNPs within a gene}
\label{tab:tab1}
\begin{tabular}{c|*{17}{c}}
\hline
\textbf{\#SNPs/gene} & 1 & 2 & 3 & 4 & 5 & 6 & 7 & 8 & 9 & 10 & 11 & 12 & 13 & 14 & 15 & 23 & 45 \\
\hline
\textbf{Count}       & 4281 & 1568 & 566 & 253 & 110 & 53 & 25 & 18 & 13 & 3 & 5 & 3 & 4 & 4 & 1 & 2 & 1 \\
\hline
\end{tabular}
\label{tab:snps_per_gene}
\end{table}

\begin{figure}[!htbp]
    \centering
    \includegraphics[width=\linewidth]{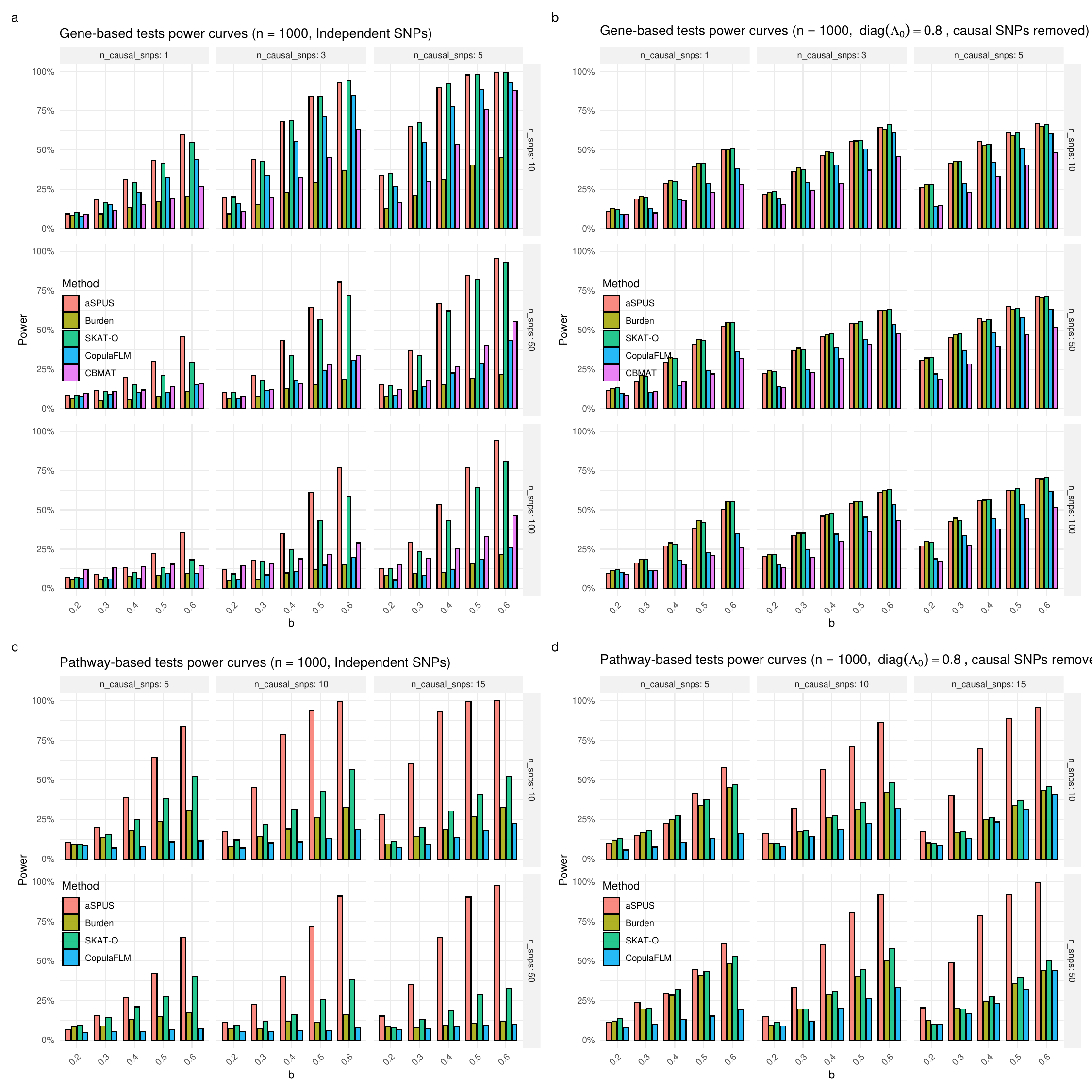}
    \caption{Simulated power for gene- and pathway-based tests}
    \label{fig:supp_power}
\end{figure}

\begin{table}[htbp]
    \centering
    \caption{Prognostic associations of selected genes across cancer types based on The Human Protein Atlas (Cancer Atlas).}
    \label{tab:cancer}
    \begin{tabular}{llll}
    \hline
    \textbf{Gene} & \textbf{Cancer Type (TCGA)} & \textbf{Prognostic Direction} & \textbf{Evidence Level} \\
    \hline
    NEB      & Kidney renal clear cell carcinoma (KIRC) & Favorable   & Potential \\
    NEB      & Lung adenocarcinoma (LUAD)               & Unfavorable & Potential \\
    BOD1L1   & Kidney renal clear cell carcinoma (KIRC) & Favorable   & Validated \\
    BOD1L1   & Liver hepatocellular carcinoma (LIHC)    & Unfavorable & Potential \\
    FICD     & Pancreatic adenocarcinoma (PAAD)         & Favorable   & Potential \\
    FICD     & Liver hepatocellular carcinoma (LIHC)    & Unfavorable & Potential \\
    PIWIL4   & Uterine corpus endometrial carcinoma (UCEC) & Unfavorable & Potential \\
    PIWIL4   & Kidney renal clear cell carcinoma (KIRC) & Unfavorable & Potential \\
    PRDX5    & Liver hepatocellular carcinoma (LIHC)    & Unfavorable & Potential \\
    PRDX5    & Ovarian serous cystadenocarcinoma (OV)   & Favorable   & Potential \\
    PRDX5    & Lung adenocarcinoma (LUAD)               & Unfavorable & Potential \\
    CYYR1    & Kidney renal clear cell carcinoma (KIRC) & Favorable   & Potential \\
    CYYR1    & Lung adenocarcinoma (LUAD)               & Unfavorable & Potential \\
    TDRD3    & Uterine corpus endometrial carcinoma (UCEC) & Favorable   & Potential \\
    TDRD3    & Liver hepatocellular carcinoma (LIHC)    & Unfavorable & Potential \\
    TMEM72   & Kidney renal clear cell carcinoma (KIRC) & Favorable   & Potential \\
    ZNF761   & Kidney chromophobe (KICH)                & Unfavorable & Potential \\
    ZNF761   & Liver hepatocellular carcinoma (LIHC)    & Unfavorable & Potential \\
    \hline
    \end{tabular}
    \label{tab:HPA_prognostic_multi}
\end{table}
\end{document}